\documentclass[%
superscriptaddress,
preprint,
 amsmath,amssymb,
 aps,
pre,
]{revtex4-1}

\usepackage{graphicx}
\usepackage{dcolumn}

\usepackage{xcolor}
\usepackage{bm}

\usepackage[colorlinks=true,linkcolor=blue,citecolor=blue]{hyperref}

\begin{document}


\title{Mutation and SARS-CoV-2 strain competition under vaccination in a modified SIR model}


\author{\textcopyright M. Ahumada}
\affiliation{Departamento de F\'isica, Universidad T\'ecnica Federico Santa Mar\'ia, Casilla 110 V, Valpara\'iso, Chile.}
\author{A. Ledesma-Araujo}
\author{Leonardo Gordillo}
\author{Juan F. Mar\'in}%
\email{juan.marin.m@usach.cl}
\affiliation{Departamento de F\'isica, Universidad de Santiago de Chile, Av. V\'ictor Jara 3493, Estaci\'on Central, Santiago, Chile.}



\date{\today}

\begin{abstract}
The crisis caused by the COVID-19 outbreak around the globe raised an increasing concern about the ongoing emergence of variants of SARS-CoV-2 that may evade the immune response provided by vaccines. New variants appear due to mutation, and as the cases accumulate, the probability of the emergence of a variant of concern increases. In this article, we propose a modified SIR model with waning immunity that captures the competition of two strain classes of an infectious disease under the effect of vaccination with a highly contagious and deadly strain class emerging from a prior strain due to mutation. When these strains compete for a limited supply of susceptible individuals, changes in the efficiency of vaccines may affect the behaviour of the disease in a non-trivial way, resulting in complex outcomes. We characterise the parameter space including intrinsic parameters of the disease, and using the vaccine efficiencies as control variables. We find different types of transcritical bifurcations between endemic fixed points and a disease-free equilibrium and identify a region of strain competition where the two strain classes coexist during a transient period. We show that a strain can be extinguished either due to strain competition or vaccination, and we obtain the critical values of the efficiency of vaccines to eradicate the disease. Numerical studies using parameters estimated from publicly reported data agree with our theoretical results. Our mathematical model could be a tool to assess quantitatively the vaccination policies of competing and emerging strains using the dynamics in epidemics of infectious diseases.
\end{abstract}

\maketitle


\section{Introduction}
\label{Sec:Introduction}

The global outbreak of the COVID-19 disease has become a major challenge for multidisciplinary scientific research, even though it is not as lethal as other diseases. While most symptoms among the population are mild, the new severe acute respiratory syndrome coronavirus 2 (SARS-CoV-2) can cause life-threatening pneumonia among some patients. Furthermore, variations in the viral strain contribute to disease severity and spreading efficiency. As a response, harsh periodical and social isolation was implemented in many countries with severe economic disruptions as a consequence \cite{Keogh2020, Aucouturier2021}.
Under this scenario, vaccination emerge as the ultimate solution to release the population.

Only a few months after the first reported case of COVID-19 in Wuhan, China (December 2019)\cite{Wei2020}, independent variants of SARS-CoV-2 were reported, such as Alpha (B.1.1.7) \cite{Tang2020}, Beta (B.1.351) \cite{Tang2021}, Gamma (P.1) \cite{Voloche2021}, Delta (B.1.617), and Omicron (B.1.1.529) \cite{GisaidDatabase}. There is thus a rising concern about the effectiveness of the currently developed vaccines in pandemic control \cite{Callaway2022}. Since SARS-CoV-2 virus attaches to human ACE2 cell surface via its surface S protein, most vaccines were developed to stimulate responses that target only the S protein of the virus \cite{Zahradnik2021}. However, mutations are causing amino acid alterations in the S protein that significantly increase the virus effectiveness to bind to the human receptors \cite{Wang2020}. According to preliminary studies, this turns into new strains that may spread more rapidly and efficiently, with consequent human losses. This has led to concerns that new strains may escape the immune response generated after vaccination \cite{Wibmer2021}.

Kermack and McKendrick proposed in their pioneering work a simplified mathematical model for the evolution of an epidemic \cite{Kermack1927}, widely known as the \emph{SIR model}. The model divides the population into three classes: \textbf{S}usceptibles, \textbf{I}nfected and \textbf{R}ecovered (SIR). A set of nonlinear differential equations govern the time evolution of the number of individuals within each class. Since the \emph{SIR} model gives a good description of the Bombay plague of 1906 \cite{Kermack1927}, many variations have been proposed to adapt the model to other infectious diseases, such as influenza, HIV, measles, malaria, SARS, foot-and-mouth disease, and whooping cough \cite{Keeling2011}. In the context of the ongoing COVID-19 pandemics, several adaptations of the SIR model have been proposed in the last two years to understand and predict the behaviour of this new coronavirus \cite{Fanelli2020, Ndairou2020, Cooper2020, Sarkar2020}. Mathematical modelling of multi-strain dynamics \cite{Keeling2011} and different approaches aiming to guide public health policies and strategies on COVID-19 vaccination \cite{Wagner2022} have lately achieved significant progress in many aspects. However, any model or approach has filled the gap of singly combining the effects of strain competition and vaccination with ongoing mutations in a coherent mathematical model, although evidence of their interplay is compelling.

In this article, we study the effect of virus mutation and vaccination in populations with two competing classes of SARS-CoV-2 strains.
We use a modified SIR model that assumes the existence of two classes of strains, one more contagious and slightly deadly than the other one. We consider the effect of vaccination in the model, assuming that vaccines have different efficiencies against each virus species. We contrasted our theoretical results with numerical simulations, with parameter values estimated from public datasets on the COVID-19 pandemics in Chile, one of the pioneering countries in massive vaccination campaigns against SARS-CoV-2 \cite{Shepherd2021, Mallapaty2021, OurWorldInData}.

The article is organised as follows: In Section \ref{Sec:SIRModel}, we introduce our modified SIR model and discuss the epidemiological assumptions made for this study. In Section \ref{Sec:FixedPoints}, we study the bifurcations of the system, as well as the equilibria and their stability. We demonstrate that the new terms in our SIR model accounting for strain competition and mutation captures a transient regime during which the stronger strains extinguish the weaker ones. In Section \ref{Sec:Numerical}, we show and discuss the results from numerical simulations of the time evolution of the pandemic. Our final remarks and conclusions are presented in Section \ref{Sec:Conclusions}.

\section{The modified SIR model for strain competition}
\label{Sec:SIRModel}

\begin{figure}
\centering
\includegraphics[width=0.3\columnwidth]{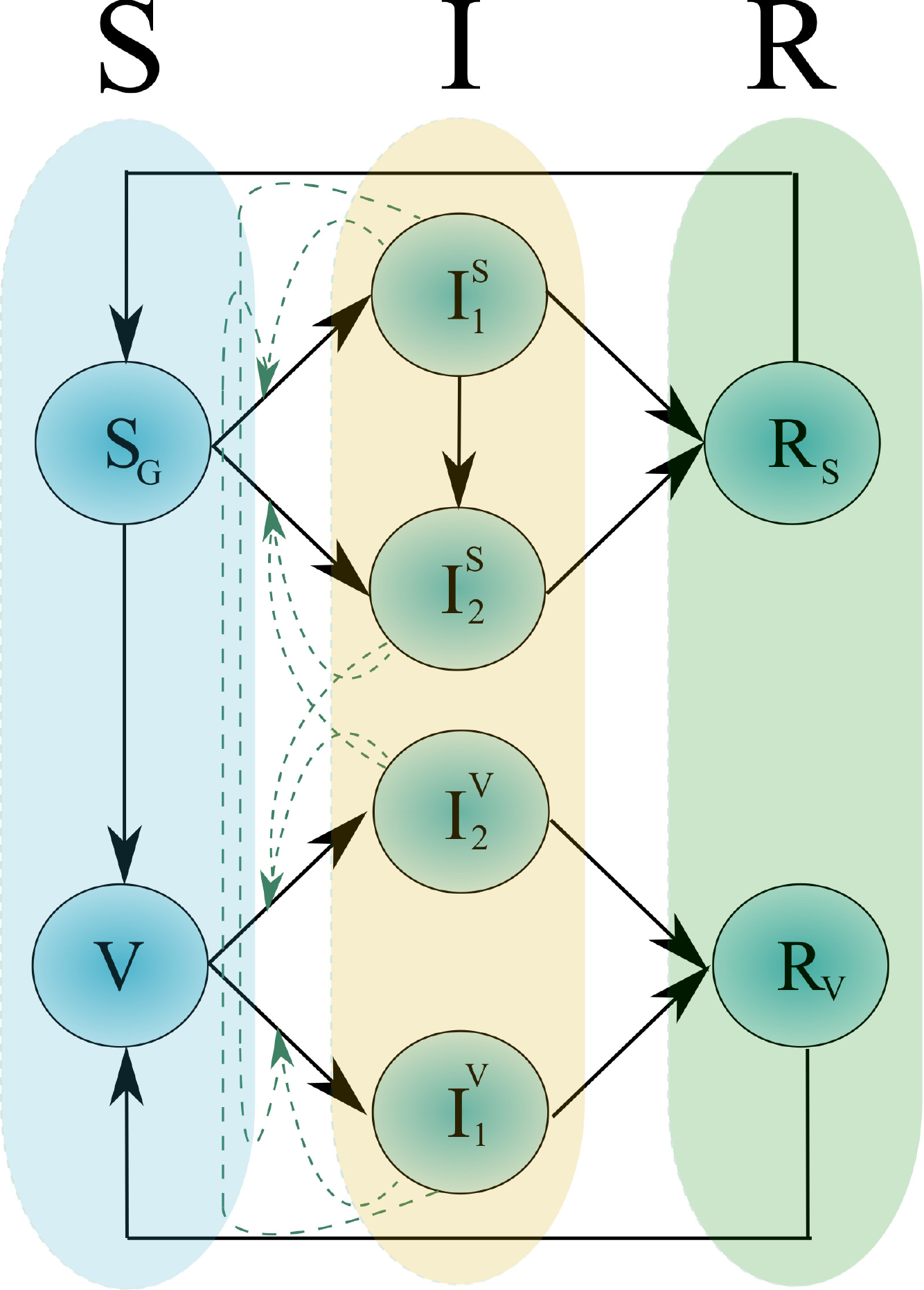}
\caption{Flow diagram of the modified SIR model summarising the main epidemiological assumptions of this study. The circulation of two strain classes of the virus is tracked among vaccinated and non-vaccinated individuals.}
\label{fig:01}
\end{figure}

The modified epidemiological SIR model schematised in Fig.~\ref{fig:01} accounts for infections for which there is no permanent immunity \cite{Keeling2011}, as it is the case of COVID-19 \cite{Altmann2021, Suryawanshi2022}. This model implicitly assumes an acute infection, i.e. a ``fast" infection where a relatively rapid immune response removes pathogens after days or weeks. Our model assumes an infectious disease spreading through one host species. Competition between strains is fundamental to disease evolution, and the scientific community has recently begun to understand the range of complex outcomes when multiple strains compete for a limited supply of susceptible individuals. Ir our model, we classify all possible strains of SARS-CoV-2 into two classes:
\begin{itemize}
    \item \textbf{Strain class 1 (SC1):} composed of those strains with moderate contagiousness and relatively low lethality. These strains can be usually controlled by vaccination with relatively high efficiency. 
    
    \item \textbf{Strain class 2 (SC2):} composed of those strains that evade the immune response provided by vaccines. These strains are more contagious and deadly than strains in class 1.
\end{itemize}

Among all possible mutations, we will focus on those of strain class 1 that turns them into a strain of class 2. That is, we are only interested in mutations that increase the contagiousness and the risks of dying from the disease.

As shown in Fig.~\ref{fig:01}, the S-class, composed of susceptible individuals, is subdivided into two classes: $\mbox{S}_{\mbox{\tiny G}}$ and V. Unvaccinated individuals are in the $\mbox{S}_{\mbox{\tiny G}}$-pool. Susceptible individuals who receive vaccination are removed from the $\mbox{S}_{\mbox{\tiny G}}$-pool and added to the vaccinated pool, the V-class. Individuals from the V-pool are not entirely immune to strain class 1 and can still be infected by strain class 2. The transition $\mbox{S}_{\mbox{\tiny G}}\to \mbox{I}_{\mbox{\tiny i}}^{\mbox{\tiny S}}\,(i=1,2)$ denotes susceptible individuals infected by the $i$-th strain class. The transition $\mbox{I}_{\mbox{\tiny 1}}^{\mbox{\tiny S}}\to \mbox{I}_{\mbox{\tiny 2}}^{\mbox{\tiny S}}$ can occur due to mutations within $\mbox{I}_{\mbox{\tiny 1}}^S$-individuals. Following the known behaviour of actual SARS-CoV-2, we neglect the possibility of an individual being infected simultaneously by the two strain classes. Thus, we assume that mutation turns $\mbox{I}^{\mbox{\tiny S}}_{\mbox{\tiny 1}}$-individuals into $\mbox{I}^{\mbox{\tiny S}}_{\mbox{\tiny 2}}$-individuals. The $\mbox{I}_{i}^{\mbox{\tiny V}}$-class ($i=1,2$) corresponds to vaccinated individuals infected by the $i$-th strain class. The complete class of infected individuals is $\mbox{I}=\mbox{I}_{\mbox{\tiny 1}}^{\mbox{\tiny S}}\cup \mbox{I}_{\mbox{\tiny 2}}^{\mbox{\tiny S}}\cup\mbox{I}_{\mbox{\tiny 1}}^{\mbox{\tiny V}}\cup\mbox{I}_{\mbox{\tiny 2}}^{\mbox{\tiny V}}$. Class $\mbox{R}_{\mbox{\tiny S}}$ ($\mbox{R}_{\mbox{\tiny V}}$) comprises unvaccinated (vaccinated) individuals recovered from the disease caused by any of the two strain classes. Infected individuals can recover after an infectious period, which is given by the transitions $\mbox{I}_{\mbox{\tiny i}}^{\mbox{\tiny V}}\to\mbox{R}_{\mbox{\tiny V}}$ and $\mbox{I}_{\mbox{\tiny i}}^{\mbox{\tiny S}}\to\mbox{R}_{\mbox{\tiny S}}$ ($i=1,2$). The complete class of recovered individuals is $\mbox{R}=\mbox{R}_{\mbox{\tiny S}}\cup \mbox{R}_{\mbox{\tiny V}}$. Eventually, $\mbox{R}$-individuals return to the $\mbox{S}$-pool due to non-permanent immunity. 

The level of the infectious disease influences the rate at which $\mbox{S}_{\mbox{\tiny G}}$($\mbox{V}$)-individuals move into any of the $\mbox{I}_{\mbox{\tiny i}}^{\mbox{\tiny S}}$($\mbox{I}_{\mbox{\tiny i}}^{\mbox{\tiny V}}$)-classes $(i=1,2)$, as indicated by the dotted arrows in Fig.~\ref{fig:01}. For the transition to any of the recovered classes, we consider a recovery rate $\gamma_i\,(i=1,2)$ for each strain class. Infections occurring in a small time interval $\mbox{d}t$,
under the hypothesis that underlying epidemiological probabilities are constant are thus modelled by the following nonlinear system of ODEs,
\begin{subequations}
\label{Eq:01}
    \begin{align}
    \label{Eq:01a}
    \dot{S}_G&=\nu+wR_S-dS_G- \upsilon\Theta(S_G)-\frac{1}{N}\left(\beta_1I_1+\beta_2I_2\right)S_G,
    \\
    \label{Eq:01b}
    \dot{V}&=\upsilon\Theta(S_G)+wR_V-dV-\frac{1}{N}\left(\bar\beta_1I_1+\bar\beta_2I_2\right)V,
    \\
    \label{Eq:01c}
    \dot{I}_1^S&=-\delta_1I_1^S-\mu I_1^S+\frac{1}{N}\beta_1 I_1S_G,
    \\
    \label{Eq:01d}
    \dot{I}_{2}^S&=-\delta_2I_2^S+\mu I_1^S+\frac{1}{N}\beta_2 S_GI_2,
    \\
    \label{Eq:01e}
    \dot{I}_1^V&=-\delta_1I_1^V+\frac{1}{N}\bar\beta_1I_1V,
    \\
    \label{Eq:01f}
    \dot{I}_2^V&=-\delta_2I_2^V+\frac{1}{N}\bar\beta_2I_2V,
    \\
    \label{Eq:01g}
    \dot{R}_S&=\gamma_1 I_1^S+\gamma_2 I_2^S-dR_S-wR_S,\\
    \label{Eq:01h}
    \dot{R}_V&=\gamma_1 I_1^V+\gamma_2 I_2^V-dR_V-wR_V,
    \end{align}
\end{subequations}
where $S_G$, $V$, $I^{S,V}_i\,(i=1,2)$, and $R_{S,V}$ are the number of individuals within their respective classes, and dots denote derivatives with respect to time. The total number of infected from the $i$-th strain class is $I_i=I_i^S+I_i^V$, with $i=1,2$. The total (time-dependent) population size is $N=S+I+R$, with $S=S_G+V$, $I=I_1+I_2$ and $R=R_S+R_V$.

The effects of demographic processes in the populations are considered with parameters $\nu$ and $d$, which are the rate at which individuals in any epidemiological class are incorporated (e.g. by births or immigration) or removed (e.g. by emigration or death by causes independent of the disease). The term $\upsilon\Theta(S_G)$ considers the vaccination, where $\upsilon$ is the vaccination rate and $\Theta$ is the Heaviside step function. Notice that the vaccination campaign is maintained only if $S_G\neq0$, i.e., as long as there are still non-vaccinated individuals. If the entire population gets vaccinated at some time $T_V$, then $\upsilon=\nu$ for $t>T_V$. Thus, the vaccination rate is adjusted to vaccine individuals at the same rate as they are incorporated into the system.

We assume a directly transmitted pathogen, so the disease transmission modelled by the transition $S\to I_i\,(i=1,2)$ is determined by only three factors: the prevalence of the infected, the underlying population contact structure, and the probability of transmission given contact. Hence, we account for homogeneous mixing that dismiss intricate patterns of contacts. The terms proportional to $\beta_i\,(i=1,2)$ in Eq.~\eqref{Eq:01} are the \emph{transmission terms} for each strain class, where $\beta_i$ is the transmission coefficient for the $i$-th strain class. Such terms are proportional to the product of the contact probability rates between susceptible and infected individuals \cite{Keeling2011}. We assume that vaccination provides protection against SC1 and SC2 with different efficiency. The latter assumption is included in the model through a modulated transmission parameter
\begin{equation}
    \label{Eq:02}
    \bar\beta_i=\beta_i(1-\eta_i)\quad (i=1,\,2),
\end{equation}
where $\eta_i$ is the efficiency of the vaccine against strains from the $i$-th class. The parameter $\mu$ in Eq.~\eqref{Eq:01} gives the mutation rate per capita at which the virus in $\mbox{I}_{\mbox{\tiny1}}^{\mbox{\tiny S}}$-individuals mutates to a strain from the second class. Parameter $\rho_i\in[0,\,1]\,(i=1,2)$ represents the per-capita probability of dying from the infection before recovering. The parameter $\delta_i$ in Eq.~\eqref{Eq:01} accounts for the combined effect of the recovery rate $\gamma_i$, $d$ and $\rho_i$, and is given by \cite{Keeling2011}
\begin{equation}
 \delta_i=\frac{\gamma_i+d}{1-\rho_i}\quad(i=1,2).   
\end{equation}
Parameter $w$ is the rate at which immunity is lost by recovered individuals and moves to either the $\mbox{S}_{\mbox{\tiny G}}$-class or the V-class. Table~\ref{Table:00} shows a summary of all parameters in our model.

\begin{table*}[]
    \footnotesize
    \centering
    \begin{tabular}{l l l l }
    \hline
    Name & Description & Value & Units \\
    \hline
    $\nu$ & Population growth rate & $3.3\times10^{-5}$ & $\mbox{day}^{-1}$ \\
    $d$ & Population decline rate & $1.7\times10^{-5}$  &  $\mbox{day}^{-1}$\\
    $\upsilon$ & Vaccination rate & 0.0043 & vaccination/day \\
    $\mu$ & Mutation rate & 0.003 & mutation/day \\
    $w$ & Coefficient of waning immunity & 0.0046 & day$^{-1}$\\
    $\beta_1$ & Transmission coefficient for the strain-class 1 & 0.2751 & day$^{-1}$ \\
    $\beta_2$ & Transmission coefficient for the strain-class 2  & 0.6641 & day$^{-1}$ \\
    $\gamma_1$ & Recovery rate for the strain-class 1 & 0.15 & day$^{-1}$ \\
    $\gamma_2$ & Recovery rate for the strain-class 2 & 0.05 & day$^{-1}$ \\
    $\rho_1$ & Probability of dying due to infection with the strain-class 1 & 0.013 & dimensionless\\
    $\rho_2$ & Probability of dying due to infection with the strain-class 2 & 0.028 & dimensionless\\
    $\eta_1$ & Efficiency of vaccines against the strain-class 1 & variable & dimensionless \\
    $\eta_2$ & Efficiency of vaccines against the strain-class 2 & variable & dimensionless\\
    \hline
    \end{tabular}
    \caption{Parameters of the modified SIR model~\eqref{Eq:01} estimated from public data-sets \cite{OurWorldInData} (see \ref{Sec:Parameterization}).
\label{Table:01}}
    \label{Table:00}
\end{table*}

Notice that our model does not consider a latent period, usually present in SEIR-type models. However, it is known that the dynamic properties of the SEIR model are qualitatively similar to those of the SIR model \cite{Keeling2011}. It is also known that there are asymptomatic cases of COVID-19 where it is unknown how much time infected individuals are transmitting the disease. The estimation from clinical data of the average time of such a latent period is controversial. Therefore, for this study, we assume the worst scenario where the infectious period begins instantly after exposure. It is also generally important to model memory effects as accurate as possible when dealing with public health issues such as the control of SARS-CoV-2. However, it is known that the qualitative dynamics are comparable in models with and without memory \cite{Keeling2011}. Therefore, only quantitative differences could be important to public health planning in a specific population. In this work, we will not consider memory effects on the behaviour of the infection since we are interested in showing the qualitative behaviour of the system under different vaccination efficiencies.

\section{Endemic and disease-free equilibria: stability analysis and bifurcations\label{Sec:FixedPoints}}

\subsection{Fixed points}

We determine the long-term behaviour of the dynamical system \eqref{Eq:01} by calculating its fixed points and stability. System \eqref{Eq:01} has a \emph{disease-free} (DF) \emph{fixed point} given by
\begin{equation}
    \label{Eq:04}
    \left(S^*_f,\,I^*_f,\,R^*_f\right)=\left(V^*_f,\,0,\,0\right)=\left(\frac{\nu}{d},\,0,\,0\right),
\end{equation}
where all the populations are zero except for the vaccinated class. There are also two \emph{endemic fixed points}: one endemic equilibrium for strain class 1 (EE1) and another endemic equilibrium for strain class 2 (EE2), given by
\begin{equation}
\begin{split}
\label{Eq:05}
    &\left(S_i^*,\,I_i^*,\,R_i^*\right)=\left(V_{(i)}^*,\,{I}_{i}^{V*},\,{R}_{V(i)}^{*}\right)\\
    &=\left(\delta_i\left(d+\gamma_i+w\right)\Lambda_i\,,\,\left(\bar\beta_i-\delta_i\right)\left(d+w\right)\Lambda_i\,,\,\left(\bar\beta_i-\delta_i\right)\Lambda_i\right),
\end{split}
\end{equation}
where $\Lambda_i=\nu/\left[\delta_i\left(d-\delta_i+\gamma_i\right)\left(d+w\right)+d\delta_i\bar\beta_i+\bar\beta_i w\left(\delta_i-\gamma_i\right)\right]$ with $i=1,2$.
To eradicate the disease, the parameters of the system must be tuned to have a stable DF fixed point. On the other hand, if the endemic fixed point for the $i$-th strain class is stable, then ${I}_i^{V}$ approaches asymptotically to a constant value in time, and the disease remains bounded in numbers within the population. In this latter scenario, ${I}_i^{V*}$ must be as small as possible to bound the accumulated number of deaths caused by the disease and avoid the collapse of hospitals and the economic impact due to isolation and quarantines. Also, Eq.~\eqref{Eq:05} shows that ${I}_i^{V*}$ is proportional to the difference $\bar\beta_i-\delta_i$. Hence, from Eq.~\eqref{Eq:02}, we conclude that it is convenient to tune $\eta_i$ to decrease ${I}_i^{V*}$ as much as possible, thus keeping the infection under control.

%

One immediate conclusion that can be drawn from Eq.~\eqref{Eq:05} is that the two strain classes cannot coexist within the same population in the long term. There are two fixed points that clearly separate the two strain classes. Thus, the dynamical system naturally selects the lasting strain as the population approaches an endemic equilibrium, where one of the strains becomes extinct. If a strain is more contagious and deadly than others and is also more resistant to vaccines, it will extinguish other weaker strains during a transient period and remain endemic in the population in the long term. This behaviour has already been reported in genomic-sequence studies of COVID-19 cases worldwide. For instance, in Chile, the ancestral strain with the Spike-D614G protein mutation was the dominant variant in sequenced cases between March 2 and April 5, 2020 \cite{Castillo2020}. The Delta variant was detected in October 2020 in India and became dominant worldwide in November 2021, conforming more than 99\% of the total cases of SARS-CoV-2 \cite{WHO_November2021}. Nowadays, the ancestral strain is almost extinct worldwide, and the circulation of the Delta strain has significantly decreased after competition with the Omicron variants. Now the Omicron are the only circulating strains categorised as variants of concern by the WHO \cite{WHO_June2022}.

Individuals within the non-vaccinated categories vanish as one approaches one of the endemic fixed points. Notice that our model does not consider the effects of anti-vax attitudes in individuals within the $S_G$ pool. Thus, there are two possible fates for the non-vaccinated population: i) they are vaccinated and thus contribute to the transition $S_G\to V$, or ii) they die due to the infection or by other independent causes. As we will show later, the system can approach an endemic fixed point in a time of the order of months, so most of the deaths before approaching one of the endemic fixed points are due to infection rather than other causes. 


\subsection{Stability of fixed points, basic reproductive ratios and bifurcations}

The linear stability analysis performed after linearising the system \eqref{Eq:01} around fixed points is algebraically challenging. We overcame the problem by rewriting Eq.~\eqref{Eq:01} as a reduced model. As we show in the following, the reduced model is equivalent to the original system if some conditions are provided. From Eq.~\eqref{Eq:01}, the global dynamics of the $S$, $I$ and $R$ populations are governed by
\begin{subequations}
   \label{Eq:06}
   \begin{align}
    \label{Eq:06a}
    \dot{S}&=\nu+wR-dS-\frac{1}{N}\left[\beta_1\left(S-\eta_1V\right)I_1+\beta_2\left(S-\eta_2V\right)I_2\right],\\
    \label{Eq:06b}
    \dot{I}&=-\delta_1I_1-\delta_2I_2+\frac{1}{N}\left[\beta_1\left(S-\eta_1V\right)I_1+\beta_2\left(S-\eta_2V\right)I_2\right],\\
    \label{Eq:06c}
    \dot{R}&=\gamma_1I_1+\gamma_2I_2-\left(d+w\right)R.
   \end{align}
\end{subequations}
We consider two cases where each endemic equilibrium can collide with the DF point via transcritical bifurcations.

\subsubsection{Case 1: Endemic equilibrium 2 bifurcating transcritically with the disease-free equilibrium\label{Subsec:Tau}}

First, we assume that SC2 is much more contagious than SC1, i.e. $\beta_2\gg\beta_1$. Assuming that $N\sim S_G\sim V$ and $\eta_1\sim\eta_2$, Eqs.~\eqref{Eq:06a} and \eqref{Eq:06b} reduce to
\begin{subequations}
   \label{Eq:07}
   \begin{align}
    \label{Eq:07a}
    \dot{S}&=\nu+wR-dS-\frac{1}{N}\beta_2\left(S-\eta_2V\right)I_2,\\
    \label{Eq:07b}
    \dot{I}&=-\delta_1I_1-\delta_2I_2+\frac{1}{N}\beta_2\left(S-\eta_2V\right)I_2.
   \end{align}
\end{subequations}
If SC2 is more deadly or displays a shorter infectious period than SC1, then $\delta_2\gg\delta_1$ and $I_1$ evolves in a different time scale than $I_2$. Notice that the nonlinear term in Eq.~\eqref{Eq:07b} does not counteract the linear term $-\delta_1 I_1$, which gives an exponential decay of $I_1$. Thus, after an initial transient characterised by an increase of $I_1(t)$ up to some maximum value, we expect an exponential decay for $I_1(t)$ with a characteristic time given by
\begin{equation}
    \label{Eq:09}
    \tau_1=\frac{1-\rho_1}{\gamma_1+d}.
\end{equation}
Thus, under the assumptions $\beta_2\gg\beta_1$ and $\delta_2\gg\delta_1$, the SC1 decays exponentially and becomes extinct after competition with SC2 at a time approximately equal to $2\tau_1$. Notice from Eq.~\eqref{Eq:09} that the characteristic time $\tau_1$ is small if individuals infected with the SC1 have a high probability of dying from the disease or have a short infectious period. This remark is in accordance with the intuitive idea that successful mutations of a strain will try to increase their contagious period as much as possible or decrease their lethality to avoid a fast extinction.

\begin{figure}
    \centering
    \scalebox{0.28}{\includegraphics{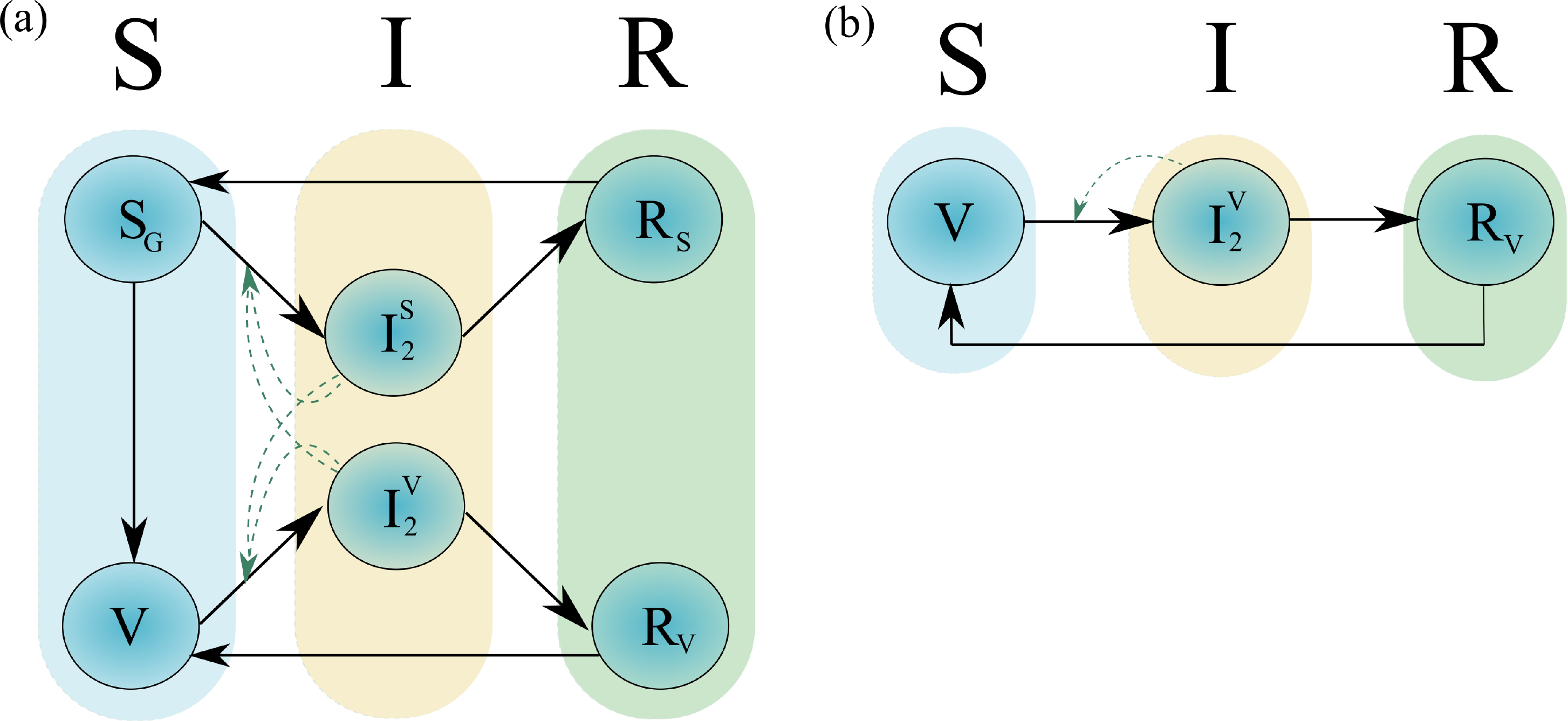}}
    \caption{Reduced SIR models capturing the dynamics of the original system of Eq.~\eqref{Eq:01} for $\beta_2\gg\beta_1$, $\delta_2\gg\delta_1$, and \textbf{(a)} $2\tau_1< t<T_{\upsilon}$, following Eq.~\eqref{Eq:10}, and \textbf{(b)} $t\gg T_{\upsilon}$, following Eq.~\eqref{Eq:11}.}
    \label{fig:02}
\end{figure}

Given that $I_1(t)$ decays exponentially after a transient, the contribution $\gamma_1I_1$ in Eq.~\eqref{Eq:06c} is negligible for $t\gg2\tau_1$. Thus, for $t\gg2\tau_1$, the system simplifies to
\begin{subequations}
   \label{Eq:10}
   \begin{align}
    \label{Eq:10a}
    \dot{S}&=\nu+wR-dS-\frac{1}{N}\beta_2\left(S-\eta_2V\right)I_2,\\
    \label{Eq:10b}
    \dot{I}&=-\delta_2I_2+\frac{1}{N}\beta_2\left(S-\eta_2V\right)I_2,\\
    \label{Eq:10c}
    \dot{R}&=\gamma_2I_2-\left(d+w\right)R,
   \end{align}
\end{subequations}
whose dynamics are represented by the reduced diagram shown in Fig.~\ref{fig:02}(a). Eventually, all individuals within the $S_G$-class will receive vaccination after some time $T_{\upsilon}$. Thus, for $t\gg T_{\upsilon}>2\tau_1$, we have $S_G\to0$, $S\to V$, and the system \eqref{Eq:10} is reduced to the following third-order dynamical system
\begin{subequations}
\label{Eq:11}
\begin{align}
    \label{Eq:11a}
    \dot{V}=&\nu+wR_V-dV-\frac{1}{N}\bar\beta_2VI_2^{V},\\
    \label{Eq:11b}
    \dot{I}_2^{(V)}=&-\delta_2I_2^{V}+\frac{1}{N}\bar\beta_2VI_2^{V},\\
    \label{Eq:11c}
    \dot{R}_V=&\gamma_2I_2^{V}-\left(d+w\right)R_V,
\end{align}
\end{subequations}
which is the conventional SIR model with demography and waning immunity \cite{Keeling2011}, represented by the flow diagram depicted in Fig.~\ref{fig:02}(b). The reduced dynamical system \eqref{Eq:11} captures the long-term asymptotic behaviour of the complete system \eqref{Eq:01} if $\beta_1\ll\beta_2$ and $\delta_1\ll\delta_2$. Indeed, Eqs.~\eqref{Eq:04} and \eqref{Eq:05} for $i=2$ are also fixed points of the reduced dynamical system \eqref{Eq:11}. In this case, the EE1 is not stable, and the SC1 becomes extinct.

With the reduced system \eqref{Eq:11}, it is possible to compute a relevant parameter: the \emph{basic reproductive ratio} for SC2, denoted here as $R_{0,\,2}$. Such parameter measures the maximum reproductive potential of infectious disease \cite{Diekmann2000} and is defined as the number of secondary cases arising from a primary case in average in an entirely susceptible population. It is also interpreted as the rate at which new cases are produced by an infectious individual multiplied by the average infectious period. Note from Eq.~\eqref{Eq:11b} that infected individuals spend an average $(1-\rho_2)/(\gamma_2+d)$ time units within the infectious class 2. Thus, the average number of new infections caused per infected individual is the transmission rate times the infection period, i.e.
\begin{equation}
    \label{Eq:12}
    R_{0,\,2}=\frac{\bar\beta_2}{\delta_2}.
\end{equation}
Performing the linear stability analysis of the system \eqref{Eq:11} around the DF equilibrium, we obtain three eigenvalues,
\begin{equation}
\label{Eq:13}
    \lambda_1=-d,\quad\lambda_2=-d-w,\quad\lambda_3=\bar\beta_2-\delta_2.
\end{equation}
Given that $d>0$ and $w>0$, from Eq.~\eqref{Eq:12} and~\eqref{Eq:13} we conclude that there is a bifurcation point at $R_{0,\,2}=1$, where the EE1 collides with the DF point in a transcritical bifurcation. If $R_{0,\,2}<1$ ($R_{0,\,2}>1$), the DF point is a stable (saddle) point.

\begin{figure}
    \centering
    \scalebox{0.5}{\includegraphics{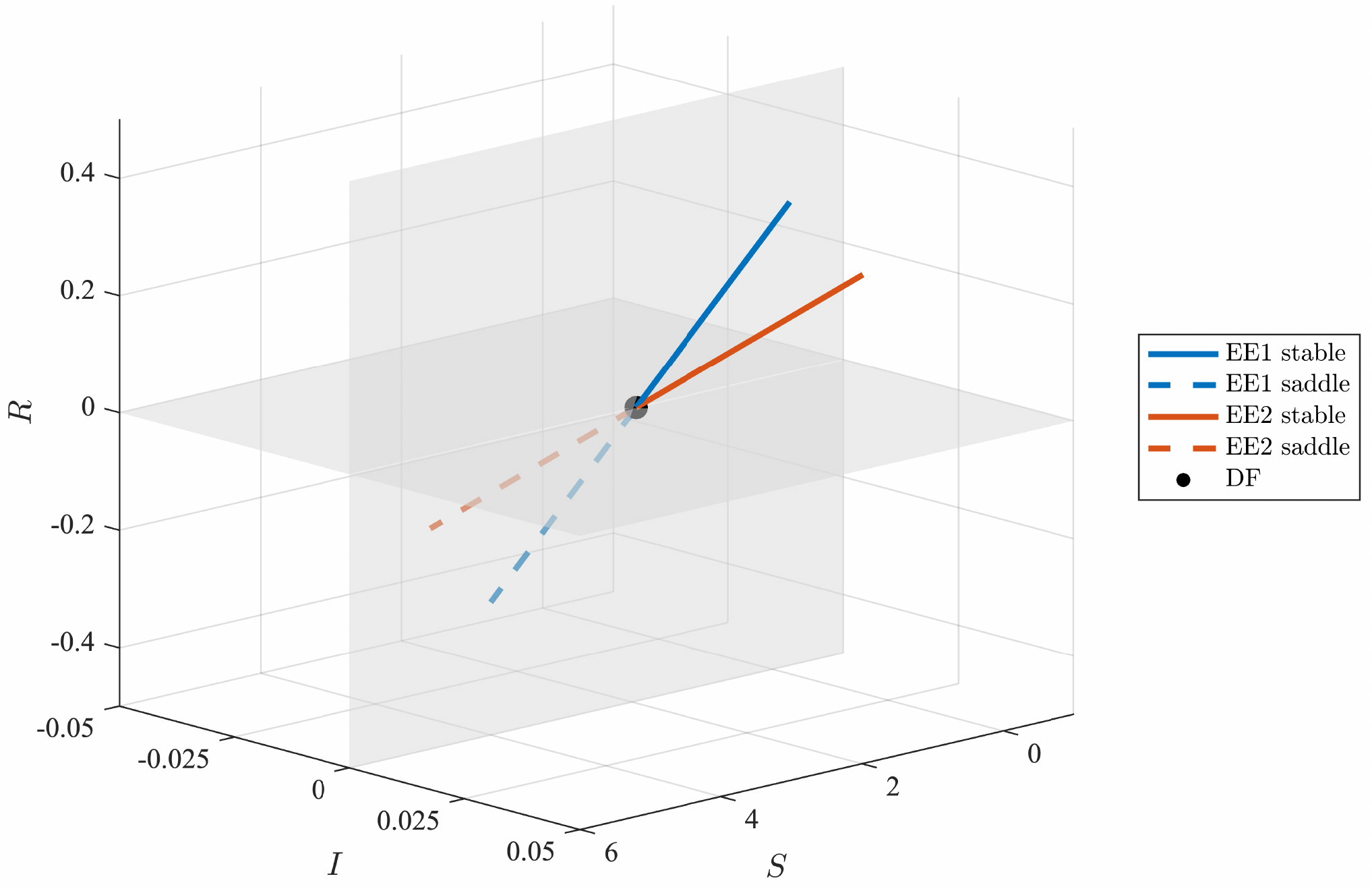}}
    \caption{Parametric curves showing the position of the equilibria of the system, with the vaccine efficiency $\eta_1$($\eta_2$) as the parameter of the EE1(EE2) curve. Transcritical bifurcations occur at $\eta_i=\eta_i^{(c)}$, where the $i$-th endemic equilibrium collides with the disease-free equilibrium ($i=1,2$). The intrinsic parameter values are shown in Table~\ref{Table:01}.}
    \label{fig:03}
\end{figure}

Figure \ref{fig:03} shows the position of the equilibria of the system in the three-dimensional $SIR$ space, as given by Eqs.~\eqref{Eq:04} and~\eqref{Eq:05}. Here, the position of the $i$-th endemic equilibrium is shown as a parametric curve of $\eta_i$, with $i=1,2$. We have complemented our theoretical results on the linear stability analysis with numerical continuation studies performed on the reduced three-dimensional system of Eq.~\eqref{Eq:11}. In Fig.~\ref{fig:03}, we show in solid (dashed) lines the points where each endemic equilibrium is stable (saddle). Notice that each endemic equilibrium can be stable only in the first octant of the space $SIR$. When $\eta_i=\eta_i^{(c)}$, where $\eta_i^{(c)}$ is some critical value, the $i$-th endemic equilibrium collide with the DF point, and their stability properties are exchanged. After the bifurcation condition, the DF remains in the same position. In contrast, the endemic point leaves the first octant and moves away from the DF equilibrium, which is a clear evidence of a transcritical bifurcation. 

\subsubsection{Case 2: Endemic equilibrium 1 bifurcating transcritically with the disease-free equilibrium}

We have shown the case where the more contagious strain SC2 extinguishes SC1. Remarkably, here we show that under certain conditions, the endemic fixed points can switch roles. If the efficiency of vaccines against SC1 decreases far below a critical value (for example, due to waning immunity from vaccines or a sudden decrease in their quality), SC2 may become extinct. The system would be characterised by an exponential decay of $I_2(t)$ and a stable endemic equilibrium for SC1. This scenario can happen even if $\beta_2$ is still greater than $\beta_1$.

Suppose in Eq.~\eqref{Eq:06} that $\eta_1<\eta_2$ such that $\bar\beta_1\gg\bar\beta_2$ and simultaneously $\beta_2\gg\beta_1$. Keeping the dominant terms in Eq.~\eqref{Eq:06} and considering that after a time $T_{\upsilon}$ all population within the $S_G$-class becomes vaccinated, Eq.~\eqref{Eq:06} reduces to
\begin{subequations}
   \label{Eq:14}
   \begin{align}
    \label{Eq:14a}
    \dot{S}&=\nu+wR-dS-\frac{1}{N}\bar\beta_1VI_1,\\
    \label{Eq:14b}
    \dot{I}&=-\delta_1I_1-\delta_2I_2+\frac{1}{N}\bar\beta_1VI_1,\\
    \label{Eq:14c}
    \dot{R}&=\gamma_1I_1+\gamma_2I_2-(d+w)R.
   \end{align}
\end{subequations}
Here, the nonlinear term in Eq.~\eqref{Eq:14b} does not counteract the linear decay term $-\delta_2I_2$. Thus, in this scenario, we expect  the extinction of the SC2, where $I_2(t)$ describes an exponential decay for $t\geq T_{\upsilon}$ with characteristic time $\tau_2=1/\delta_2$ [in analogy with Eq.~\eqref{Eq:09}]. Finally, we arrive at a completely analogous third-order system as Eq.~\eqref{Eq:11}, replacing $\bar\beta_2\to\bar\beta_1$, $\delta_2\to\delta_1$, $\gamma_2\to\gamma_1$, and $I_2^{V}\to I_1^{V}$. Equations \eqref{Eq:04} and \eqref{Eq:05} for $i=1$ are now the fixed points of the reduced dynamical system, which now captures the endemic equilibrium for the SC1. Thus, we have a basic reproductive ratio for SC1 analogous to Eq.~\eqref{Eq:12}, given by
\begin{equation}
    \label{Eq:15}
    R_{0,\,1}=\frac{\bar\beta_1}{\delta_1}.
\end{equation}
Similarly, the EE1 will bifurcate transcritically with the DF equilibrium at $R_{0,\,1}=1$.

\begin{figure*}
    \centering
    \scalebox{0.35}{\includegraphics{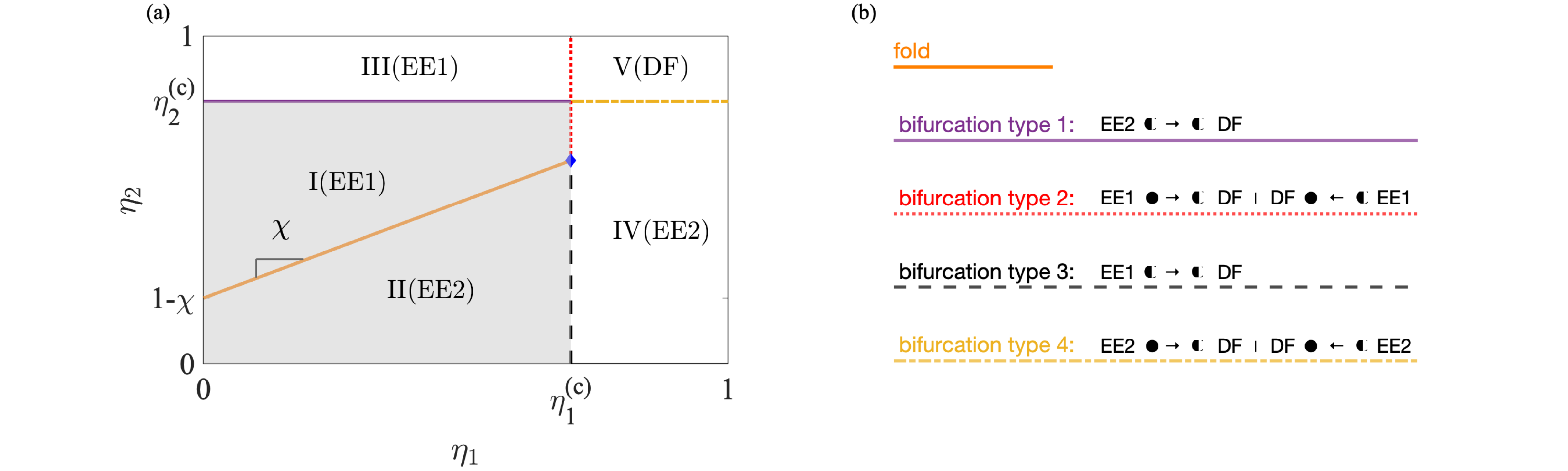}}
    \caption{\textbf{(a)} Parameter space of the system for $\chi<\chi_{\max}$. The stable fixed point in each region is indicated in parentheses (the remaining equilibria are saddle points). The shaded regions comprises the \emph{strain competition region}, where both endemic equilibria coexist and compete. Along the orange solid line of slope $\chi$, there is a fold where the endemic equilibria exchange their stability. \textbf{(b)} Diagram with the different types of observed bifurcations. One fold and four transcritical bifurcations can occur, where one endemic equilibrium collides with the DF point. Different lines indicate different bifurcations (see explanation in the text).}
    \label{fig:04}
\end{figure*}

\subsubsection{Case 3: Fold between endemic equilibria}

The reproductive ratio of a given strain class is one of the parameters determining if it will remain endemic in the population in the long term. From the results discussed above, if only one of the two reproductive ratios is above unity, the corresponding strain will extinguish the other and remain endemic within the population. Notwithstanding, there is an intermediate situation where both reproductive ratios are above unity. Given that the two strain classes cannot coexist within the population in an equilibrium state, both strains will compete for the limited population of susceptible individuals, and the system will approach the endemic equilibrium corresponding to the strain with the largest reproductive ratio. Indeed, both endemic fixed points coexist if $R_{0,1}>1$ and $R_{0,2}>1$, being stable (saddle) the one with the largest (smallest) reproductive ratio. A fold occurs at $R_{0,1}=R_{0,2}$, where both fixed points exchange their stability properties.

\subsection{Bifurcations and strain competition}

In previous sections, we have shown that by changing parameters $\eta_1$ and $\eta_2$, each endemic equilibrium can change its stability and collide independently with the DF equilibrium via a transcritical bifurcation. The bifurcation point for each strain determines a critical value of the vaccine efficiency against the corresponding strain, given by
\begin{equation}
\label{Eq:17}
    \eta_i^{(c)}=1-\frac{\gamma_i+d}{\beta_i(1-\rho_i)},\quad i=1,2.
\end{equation}
Suppose that $\eta_i<\eta_i^{(c)}$ with $i=1,2$. As it has been discussed, strain competition leads to the extinction of the strain with the lowest reproductive ratio. After that, increasing $\eta_i$ above the critical value $\eta_i^{(c)}$, where $i$ denotes the surviving strain with the largest reproductive ratio, collides the associated endemic equilibrium with the DF equilibrium, and consequently exchanging their stability properties via a transcritical bifurcation. Moreover, the $i$-th endemic equilibrium becomes inaccessible to the system after the bifurcation since at least one of its coordinates becomes negative and loses biological sense. Thus, having $\eta_i>\eta_i^{(c)}$ for both strains is favourable for the population if we wish to eradicate the disease. Note that a highly contagious and lethal strain will have large values of both $\beta_i$ and $\rho_i$, requiring a vaccine efficiency near unity to eradicate the disease. 

Figure \ref{fig:04} shows the parameter space of the system, summarising our theoretical predictions on the long-term behaviour of the dynamical system. In Fig.~\ref{fig:04}(a), we enumerate the different regions of the parameter space, which are separated by boundaries where different bifurcations can occur. The possible bifurcations are shown in the diagram of Fig.~\ref{fig:04}(b), where half-filled circles denote saddle points, and filled circles denote attractors.

In regions I and II, the endemic equilibrium with the largest (smallest) reproductive ratio is a stable (saddle) point. Such regions comprise what we labelled as the \emph{strain competition region}, and is depicted as a shaded region in Fig.~\ref{fig:04}(a). The condition $R_{0,1}=R_{0,2}$ is fulfilled along the orange solid line separating regions I and II, as shown in Fig.~\ref{fig:04}(a), whose locus are solutions to the equation
\begin{equation}
    \label{Eq:16}
    \eta_2=\chi(\eta_1-1)+1,\quad \chi=\frac{\beta_1\delta_2}{\beta_2\delta_1},
\end{equation}
where $0<\chi<\chi_{\max}$. Here, $\chi_{\max}$ is the maximum value that $\chi$ can have and is given by
\begin{equation}
\chi_{\max}=\frac{\eta_2^{(c)}-1}{\eta_1^{(c)}-1}.    
\end{equation}
Figure~\ref{fig:04}(a) shows five possible bifurcations for $\chi<\chi_{\max}$. Below we summarise the properties of the system in each region of parameter space:
\begin{itemize}
    \item \textit{Region I}: The EE1 is the attractor of the system, whereas the EE2 and the DF equilibrium are saddle points. This region is the upper part of the strain competition region, and is bounded from below by the fold line~\eqref{Eq:16} where endemic points exchange their stability properties. The region is bounded from above by the bifurcation type 1, where the saddle EE2 and the saddle DF equilibrium collide via a transcritical bifurcation. The region is also bounded from the right by the bifurcation type 2. Approaching the bifurcation type 2 from region I, the stable EE1 collides with the saddle DF point in a transcritical bifurcation.
    
    \item \textit{Region II}: The EE2 is the attractor of the system, whereas the EE1 and the DF equilibrium are saddles. This region is the lower part of the strain competition region, is bounded from above by the fold line, and from the right by the bifurcation type 3, where the saddle EE1 collides with the saddle DF equilibrium via a transcritical bifurcation.
    
    \item \textit{Region III}: The EE1 is the attractor of the system, whereas the EE2 and the DF equilibrium are saddles. The region is bounded from below by the transcritical bifurcation type 1 and from the right by the bifurcation type 2. The same bifurcation is observed approaching the bifurcation type 2 either from region I or from region III.
    
    \item \textit{Region IV}: The EE2 is the attractor of the system, whereas the EE1 and the DF equilibrium are saddles. The region is bounded from the left by the bifurcation type 3 and from above by the bifurcation type 4. Approaching the bifurcation type 4 from region IV, the stable EE2 collides with the saddle DF point via a transcritical bifurcation.
    
     \item \textit{Region V}: The DF equilibrium is the attractor, whereas the EE1 and the EE2 are saddles. This region is bounded from the left by the bifurcation type 2 and from below by the bifurcation type 4. Approaching the bifurcation type 2(4) from region V, the saddle EE1(EE2) collides with the stable DF point in a transcritical bifurcation. This region of parameter space is the most favourable for the population.
\end{itemize}

It is worth noting that a double-bifurcation point can occur if $\chi=\chi_{\max}$, where the crossing point indicated with a blue diamond in Fig.~\ref{fig:04} coincide with the crossing point between the thresholds $\eta_1=\eta_1^{(c)}$ and $\eta_2=\eta_2^{(c)}$. However, the parameters of the system should be fine-tuned to observe such a special bifurcation: the reproductive ratio of both strains must be equal, and the efficiency of vaccines must match their critical values. Moreover, although it is realistic to tune vaccine efficiencies, it is still an open question whether it is possible to alter the recovery rate of COVID-19 patients, which would be the only realistic way to change the reproductive ratio of both strain classes with $\eta_i$ fixed.  Hence, the double-bifurcation will not be observed in practical situations.

\section{Numerical simulations and continuation of stationary solutions: time evolution of cases and bifurcations\label{Sec:Numerical}}

We perform a numerical study of the dynamical system~\eqref{Eq:01} in parameter space $(\eta_1,\eta_2)$ through two methods of analysis: using \emph{numerical continuation} techniques  \cite{Allgower2012, Krauskopf2007, AUTO} and using direct numerical simulations of Eq.~\eqref{Eq:01}. We estimate the parameters of the system based on the public database on the COVID-19 pandemic from the European Centre for Disease Prevention and Control (ECDC) \cite{OurWorldInData}. In \ref{Sec:Parameterization}, we give the details on the methods for the estimation of such parameters, which are summarised in Table \ref{Table:01}. We have used these values of parameters for the parametric curves of Fig.~\ref{fig:03} and for the numerical simulations discussed below.

\subsection{Numerical continuation}

\begin{figure*}
    \centering
    \scalebox{0.49}{\includegraphics{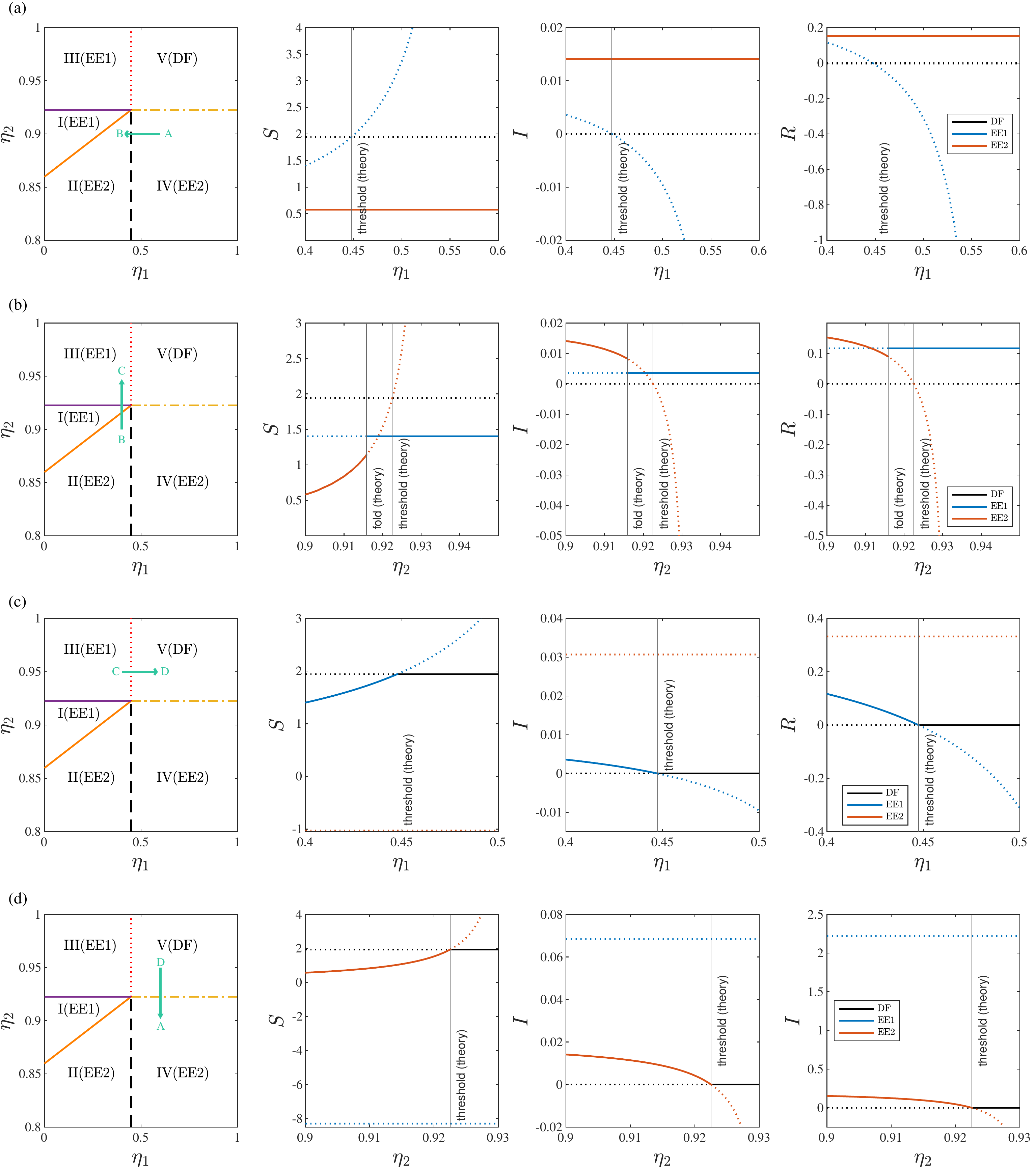}}
    \caption{Bifurcation diagrams of the dynamical system~\eqref{Eq:01}. Results obtained using numerical continuation along a closed-loop joining regions \textbf{(a)} IV and II (path $A\to B$, with $0.4\leq\eta_1\leq0.6$ and $\eta_2=0.9$), \textbf{(b)} II, I, and III (path $B\to C$, with $0.9\leq\eta_2\leq0.95$ and $\eta_1=0.4$) \textbf{(c)} III and V (path $C\to D$, with $0.4\leq\eta_1\leq0.6$ and $\eta_2=0.95$), and  \textbf{(d)} V and IV (path $D\to A$, with $\eta_1=0.6$ and $0.9\leq\eta_1\leq0.95$). The left panel show the path in parameter space. The rightmost panels show the $SIR$ components of the stationary solutions. The theoretical thresholds of the fold and transcritical bifurcations are shown with vertical solid lines. Stable (saddle) points are denoted with solid (dashed) lines. } 
    \label{fig:05}
\end{figure*}

Using numerical continuation, we analyse the stationary solutions and bifurcations of the dynamical system~\eqref{Eq:01}. The bifurcation diagrams are shown in Fig.~\ref{fig:05}. To analyse all the bifurcations, we follow a closed loop the in parameter space from A to B [Fig.~\ref{fig:05}(a)], B to C [Fig.~\ref{fig:05}(b)], C to D [Fig.~\ref{fig:05}(c)], and D to A [Fig.~\ref{fig:05}(d)]. 

\begin{itemize}
\item \textit{Path} $A\to B$: Starts in region IV and crosses to region II through the bifurcation type 3. In accordance with our predictions in Section~\ref{Sec:FixedPoints}, the EE1 and the DF equilibrium are saddle points and collide via a transcritical bifurcation. The EE2 remains stable and unchanged along this path. Notice that the $I$ and $R$ components of the EE1 become negative for $\eta_1>\eta_1^{(c)}$. Thus, the EE1 leaves the first octant of the $SIR$ space after the transcritical bifurcation, as we already noticed in our theoretical analysis in Section~\ref{Sec:FixedPoints} [see Fig.~\ref{fig:03}].

\begin{figure*}
\centering
    \scalebox{0.34}{\includegraphics{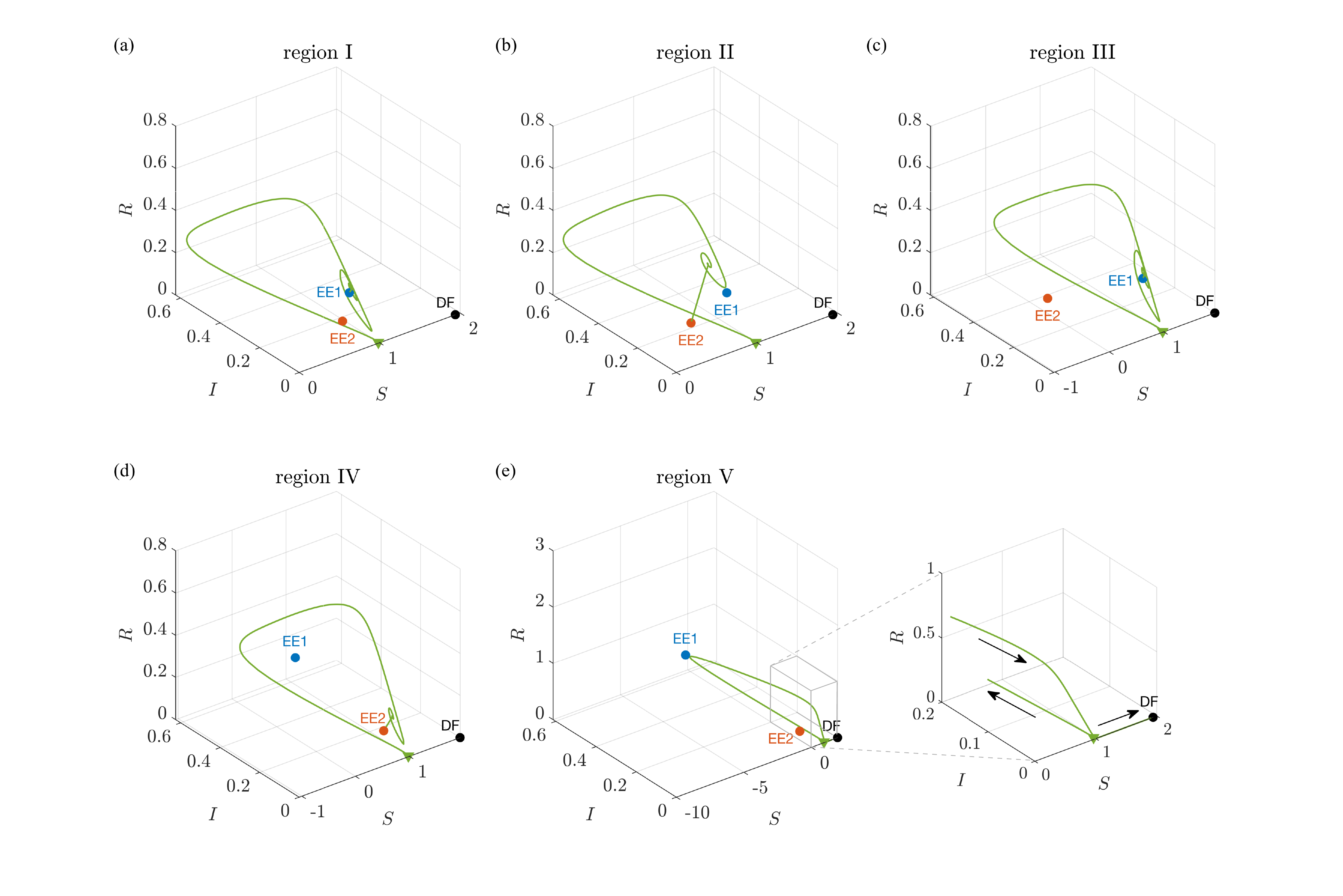}}
    \caption{Numerical simulations of the time evolution of the modified SIR model. (a) Region I: $\eta_{1}=0.20$ and $\eta_{2}=0.90$. (b) Region II: $\eta_{1}=0.20$ and $\eta_{2}=0.85$. (c) Region III: $\eta_{1}=0.20$ and $\eta_{2}=0.95$. (d) Region IV: $\eta_{1}=0.80$ and $\eta_{2}=0.90$. (e) Region V: $\eta_{1}=0.60$ and $\eta_{2}=0.97$. The parameters used are shown in Table~\ref{Table:01}. Arrows in the inset denote time direction. The initial condition for each case is indicated with a green triangle. }
    \label{fig:06}
\end{figure*}

\item \textit{Path} $B\to C$: Starts in region II, passes through the region I, and ends up in region III. Thus, the path crosses two boundary lines: the fold and the bifurcation type 1. As expected, the DF equilibrium is still a saddle point and remains unchanged along this path. Strain competition takes place in regions I and II. Since point B is in region II (below the fold condition), the EE2 is an attractor, whereas the EE1 is a saddle point. At the fold condition, the endemic equilibria exchange their stability properties, and the EE1 becomes the attractor of the system. Remarkably, the bifurcation diagrams in Fig.~\ref{fig:05}(b) confirm that such exchange occurs at distance: both points are separated by a small distance at the fold condition. Further increasing the value of $\eta_2$ above the fold condition, we observe the transcritical bifurcation type 1. Similarly, as in the case discussed above, the EE2 leaves the first octant of the $SIR$ space after the transcritical bifurcation, in accordance with Fig.~\ref{fig:03}.

\item \textit{Path} $C\to D$: Goes from region III to region V crossing the transcritical bifurcation type 2. The bifurcation diagrams shown in the rightmost panels of Fig.~\ref{fig:05}(c) confirm that the EE2 remains outside the first octant of the $SIR$ space as a saddle point all along this path. At point $C$, the EE1 is stable, whereas the DF equilibrium is a saddle point. As we approach the threshold, the EE1 approaches the DF point and collides via a transcritical bifurcation. For $\eta_1$ above the threshold, the saddle EE1 leaves the first octant of the $SIR$ space, and the DF equilibrium becomes the attractor of the system.

\item \textit{Path} $D\to A$: Goes from region V to region IV, crossing the transcritical bifurcation type 4. The EE1 remains outside the first octant of the SIR space as a saddle point. At point $D$, $\eta_2>\eta_2^{(c)}$ and the DF equilibrium is the attractor of the system, whereas the EE2 is a saddle point. Crossing the bifurcation type 4, the EE2 collides with the DF equilibrium and exchanges their stability in a transcritical bifurcation. 
\end{itemize}
\subsection{Time evolution of cases}

We also performed direct numerical simulations of Eq.~\eqref{Eq:01} to characterise the time evolution of the modified SIR model given some initial conditions. For the time integration, we used the Dormand-Prince algorithm as an explicit Runge-Kutta of orders 4 and 5 with an adaptive time step \cite{MatlabODE45}. Based on our assumptions and public-data analysis, we use at $t=0$ the values $\mbox{I}_{\tiny 1}^{\tiny\mbox{S}}=0.0042$, $S_G=0.9958$, and $V=I_1^{V}=I_2^{S}=I_2^{V}=R_{S}=R_{V}=0$ [see \ref{Sec:Parameterization}]. Figure~\ref{fig:06} shows the trajectory of the dynamical system in each of the five regions indicated in the parameter space of Fig.~\ref{fig:04}. The trajectories are shown along with the position of the endemic equilibria, EE1 and EE2, and the DF point. In each case, we indicate the corresponding region in parameter space.

Figure~\ref{fig:06}(a) shows the resulting trajectory for $(\eta_1,\,\eta_2)$ in region I. After an initial burst in the number of infections, the system approaches the EE1, where the number of infections with strain-class 1 exhibits damped oscillations around the EE1. In Figure~\ref{fig:06}(b), we cross the fold line in the parameter space of Fig.~\ref{fig:04}, and the trajectory approaches the EE2, the attractor of the system in the corresponding region. After the initial burst of infections, the trajectory exhibits a short-lived oscillation before approaching the EE2 monotonically. Figures~\ref{fig:06}(a) and~\ref{fig:06}(b) illustrate the phenomenon of strain competition, where both endemic equilibria coexist and compete for resources. The stable equilibrium corresponds to the strain with the largest reproductive ratio.

\begin{figure*}
    \centering
    \scalebox{.65}{\includegraphics{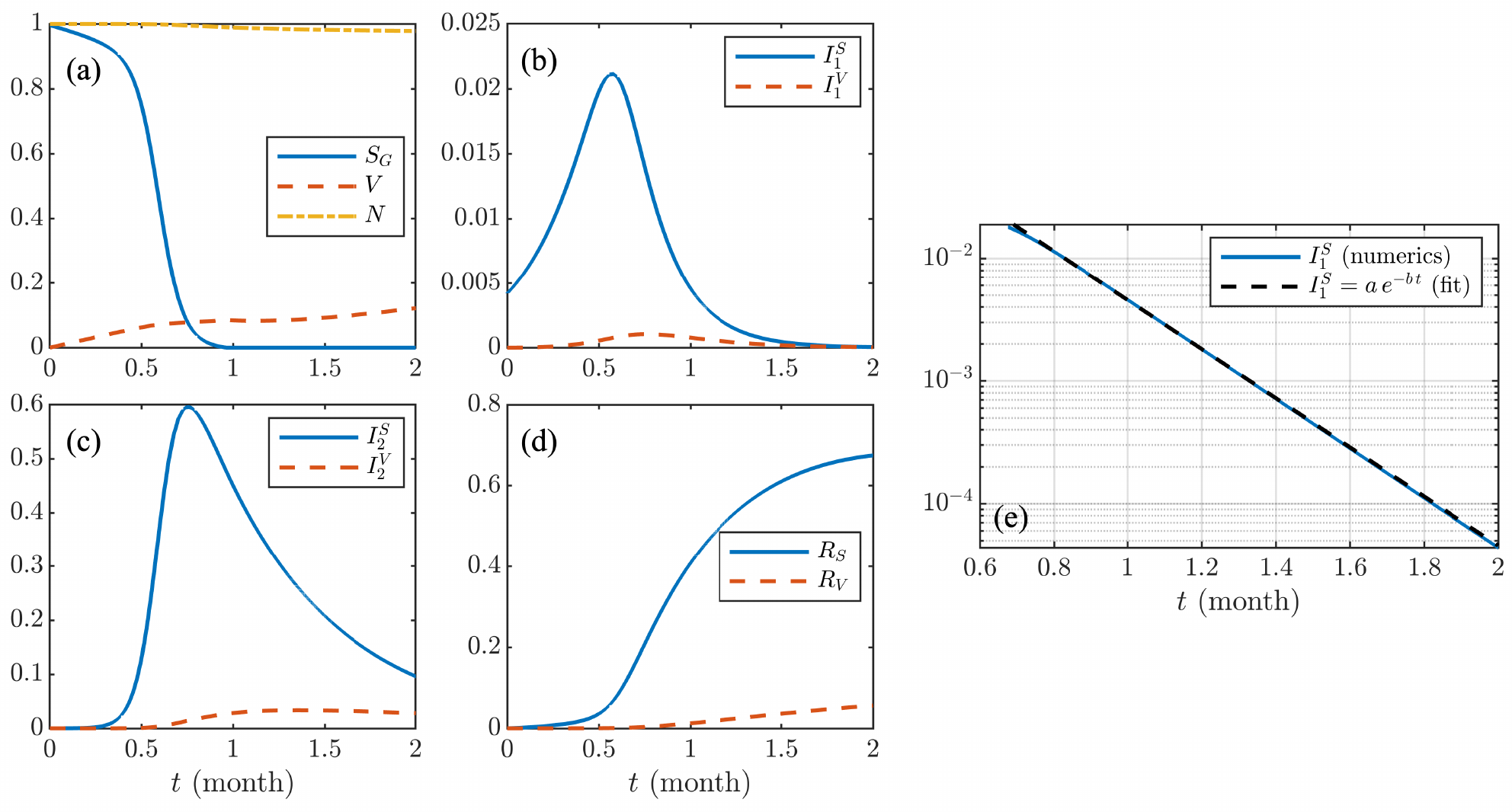}}
    \caption{Numerical simulation and characteristic decay time with $\eta_{1}=0.5$ and $\eta_{2}=0.9$, where $\eta_{1}$ and $\eta_{2}$ are in region IV. The fitted values of the parameters in panel (e) are $a=0.4567$ with $a\in[0.4561,0.4573]$, $b=4.606$ with $b\in[4.605,4.607]$, and $R$-squared 0.9998.}
    \label{fig:07}
\end{figure*}

In Fig.~\ref{fig:06}(c), the resulting trajectory is qualitatively similar to Fig.~\ref{fig:06}(a). Indeed, the EE1 is the attractor of the system in regions I and III. Similarly, the trajectory in Fig.~\ref{fig:06}(d) is qualitatively similar to Fig.~\ref{fig:06}(b). However, in contrast to regions I and II, the outcome of the system in regions III and IV are not due to strain competition. In Fig.~\ref{fig:06}(c), $\eta_2>\eta_2^{(c)}$ and the EE2 is a saddle point outside the first octant. Similarly, in Fig.~\ref{fig:06}(d), $\eta_1>\eta_1^{(c)}$ and the EE1 is outside the first octant.

Figure~\ref{fig:06}(e) shows the trajectory in region V, where the DF equilibrium is the attractor of the system. In this case, there is an initial burst in the number of infections, reaching a maximum. Later, the number of infections decreases to zero, and the trajectory passes near the initial condition of the simulation, as shown in the zoom-in of the boxed region in the inset of Fig.~\ref{fig:06}(e). After that, the trajectory goes monotonically towards the DF equilibrium. This scenario corresponds to the case where the vaccine efficiency is high against both strain classes, reducing the number of infected individuals to zero in a finite time.

Finally, we have estimated from numerical simulations the characteristic time $\tau_1$ given by Eq.~\eqref{Eq:09}. In Fig.~\ref{fig:07}, we show the time evolution of the total population $N$, $S_G$ and $V$ [Fig.~\ref{fig:07}(a)], individuals infected with strain-class 1 [Fig.~\ref{fig:07}(b)], individuals infected with strain-class 2 [Fig.~\ref{fig:07}(c)], and recovered individuals [Fig.~\ref{fig:07}(d)]. The values of the parameters correspond to region IV. Notice that the SC1 becomes extinct due to competition with SC2. The latter remains endemic in the population in the long term. This behaviour is observed in regions from I to IV, where one of the endemic equilibria is stable. In Section~\ref{Subsec:Tau}, we predicted that $I_1(t)$ decays exponentially after an initial burst, as evidenced in Fig.~\ref{fig:07}(b). In Fig.~\ref{fig:07}(e), we fit the numerical outcome for the $I_1^{S}$ individuals with an exponential decay law for $t\in[0.6,\,2]\,\mbox{month}$. We obtain $\tau_1^{\tiny fit}=0.21711\pm10^{-5} \,\mbox{month}$, which is near the theoretical value $\tau_1=0.21931\,\mbox{month}$.



\section{Conclusions}
\label{Sec:Conclusions}

Motivated by the ongoing COVID-19 pandemic caused by multiple strains of SARS-CoV-2, we pose a modified SIR model with waning immunity capturing strain competition between two classes of strains of an infectious virus under the effect of vaccination. We consider a mutation parameter that characterises the rate at which the SC1 mutates into a more deadly and contagious strain class, the SC2. We hypothesise that vaccination modulates the transmission coefficient of each strain, decreasing its value by a proportion given by their efficiency. Our model assumes that vaccination does not provide full immunity against any of the strain classes. We characterise the parameter space of the system and bifurcations using the vaccine efficiencies as control parameters. The basic reproductive ratio is determined for each strain class. We determined a region in parameter space where the disease-free equilibrium is stable, which is the most favourable scenario for eradicating the disease. We also found a region in which both strain classes coexist and compete. 

Our model shows that strain competition always leads to the extinction of one of the strain classes. In the region of strain competition, we show that after a transient period, the strain with the largest reproductive ratio remains endemic in the long term. We obtain the minimum value of the vaccine efficiency against each of the strain classes to eradicate the disease. Such critical efficiency depends on the transmission rate, the infectious period, and the probability of dying due to the infection from each strain class. We also derive an expression for the characteristic exponential decay time at which the strain extinguishes.

We conclude that the efficiency of vaccines controls the stability of the endemic equilibria. Thus, vaccines are determinant in the long-term behaviour of the pandemic. A combination of vaccine efficiencies against both strain classes could yield an endemic SC1, an endemic SC2, or even eradicate the disease. Moreover, we showed that the infectious period controls the characteristic decay time of the exponentially decaying strain. We estimated the values of the parameters of our modified SIR model based on public databases on the COVID-19 pandemic from the ECDC. Direct numerical simulations of the time evolution of the dynamical system and numerical characterisations of bifurcations using numerical continuation techniques are in good agreement with theoretical results. Our study emphasises the importance of SARS-CoV-2 genomic surveillance in regions where the virus is freely transmitted, a condition that increases the probability of emergent variants of concern due to mutation. Surveillance programmes are key in combination with a continuous improvement of vaccine technology facing the emergence of new variants of concern.

\begin{acknowledgments}
A.L.A. thanks ANID-Subdirección de Capital Humano/Doctorado Nacional/2021-21211330 for the financial support. J.F.M. acknowledge the financial support of ANID, through the grant FONDECYT/POSTDOCTORADO/3200499. This article is dedicated to the memory of Prof. Enrique Tirapegui.

\end{acknowledgments}

 \appendix

\section{Data analysis and parameterization\label{Sec:Parameterization}}

We rely on freely available data on the COVID-19 pandemic from the ECDC \cite{OurWorldInData}. The public data used in this study comprises the number of people fully vaccinated against COVID-19, the number of active cases per day, and the infection fatality rate (IFR) as a function of time. The estimate of the population growth is retrieved from the World Population Prospects 2019 Revision by the United Nations (UN) population division \cite{OurWorldInDataPopulationGrowth}, whereas the demographic parameters $\nu$ and $d$, are gathered from the World Bank collection (WBC) of development indicators \cite{WorldBank}. 

\begin{figure*}
    \centering
    \scalebox{0.27}{\includegraphics{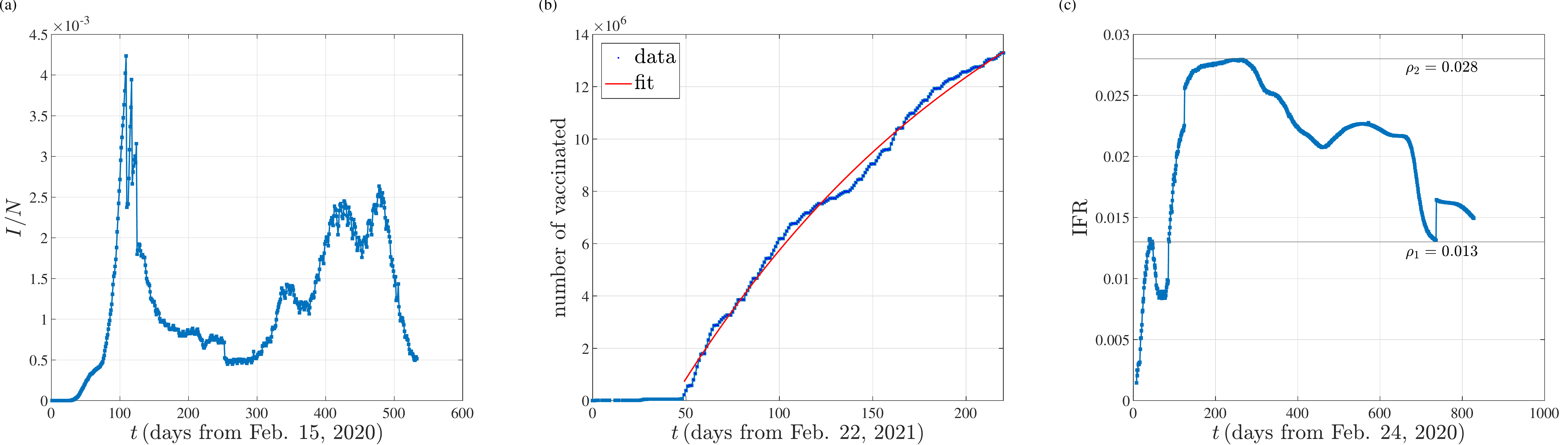}}
    \caption{Data retrieved from public databases on the COVID-19 pandemic in Chile. \textbf{(a)} Number of active cases per day. \textbf{(b)} Number of individuals who received all prescribed doses of vaccines (dotted line). The third-order polynomial fitting of the data is shown in solid red line. \textbf{(c)} The infection fatality rate as a function of time used to estimate the mortality risk of each strain class.}
    \label{fig:08}
\end{figure*}

For the initial conditions, we considered that infected individuals join the population under study at $t=0$, and at the same time, the vaccination campaign starts. Thus, $V(t=0)=$ $I_1^{V}(t=0)=$ $I_2^{S}(t=0)=I_2^{V}(t=0)=R_{S}(t=0)=R_{V}(t=0)=0$. According to UN sources, the total population in Chile by 2021 was 19.116.209 inhabitants \cite{OurWorldInDataPopulationGrowth}. This population correspond to the initial condition $N(t=0)=1$ in our numerical simulations. To estimate a realistic value for the initial infected population, $I(t=0)$, we retrieved the number of active cases as a function of time from public data \cite{Worldometers}, depicted in Fig.~\ref{fig:08}(a). We chose for $I(t=0)$ the condition corresponding to the maximum number of active cases, which occurred on June 3, 2020, with 80.933 active cases across the country. This corresponds to $I(t=0)=0.0042$, which is assigned to the sub-class $\mbox{I}_{\tiny 1}^{\tiny\mbox{S}}$ at $t=0$. This is how infections with strain-class 2 are seeded by mutation. Thus, our initials conditions are $\mbox{I}_{\tiny 1}^{\tiny\mbox{S}}(t=0)=0.0042$, $S_G(t=0)=0.9958$, with the remaining sub-classes equal to zero.

To estimate the vaccination rate $\upsilon$, we retrieved the number of people who received all doses prescribed by the vaccination protocol in Chile as a function of time. The number of vaccinated people shows a marked increase after March 2, 2021 \cite{Mathieu2021}, as shown in Fig.~\ref{fig:08}(b). We fit the data from such date on, using a third-order fitting polynomial $p_{v}(t)$, the highest order polynomial that gives a well-posed fitting of the data. Finally, we estimate the vaccination rate as $\left.\dot{p}_{v}\right|_{t=t_l}=81723\,\mbox{[people/day]}$, where $t_l$ denotes the time of lecture of the data. Normalising the vaccination rate to the total population in Chile by 2021 gives $\upsilon=0.0043\, \hbox{[vaccination/day]}$.

We estimate the value of the mutation rate $\mu$ as the inverse of the time $T_{\mu}=10\,\mbox{[month]}$ between the first reported case of COVID-19 and the first detection of the Delta variant \cite{OurWorldInData}. The first case was notified in China on December 31, 2019 \cite{Wei2020}, whereas the first case of the Delta variant was notified in India on October 2020 \cite{GisaidDatabase}. Thus, mutations leading to a VOC happens at a rate  $\mu=0.003\,\mbox{[mutation/day]}$, approximately.

Given that the population decline rate $d$ excludes the deaths due to the infection, we estimate $d$ from the number of deaths just after the COVID-19 breakout in Chile. The number of deaths in 2019 was $117\,490$ \cite{owidlifeexpectancy}. Normalising by the total population in Chile, the population decline rate is $d=1.7\times 10^{-5}\,\mbox{[1/day]}$. The birth rate in Chile in 2020 was $227\,040$ births \cite{OurWorldInDataPopulationGrowth}. Similarly, after normalisation by the total population in Chile, we obtain a population growth rate $\nu=3.3\times 10^{-5}\,\mbox{[1/day]}$.

To estimate the mortality risk of each strain class, we analysed the data of the infection fatality rate (IFR), defined as the number of deaths from the disease divided by the total number of cases. Although the total number of cases of COVID-19 is not precisely known, mainly because not every infected individual is tested \cite{Read2021}, the IFR calculated from observed data is a good estimation of how likely it is for someone infected to die from the disease \cite{OurWorldInData, Kobayashi2020}. Figure~\ref{fig:06}(c) shows the IFR as a function of time observed in Chile \cite{OurWorldInData}. We estimate the value of $\rho_2$ as the global maximum of the IFR, which was reached soon after 200 days from February 24, 2020. We estimate the value of $\rho_1$ as the minimum IFR reached after the global maximum.

We considered that immunity remained seven months after infection, based on a recent study of the risk for reinfection after SARS-CoV-2 with wild-type or Alpha variants \cite{Helfand2022}. Following the study of Byrne \textit{et al.} \cite{Byrne2020}, the duration of the infection period for the classes $\mbox{I}_{\mbox{\tiny 1}}$ and $\mbox{I}_{\mbox{\tiny 2}}$ are $T_1=6.5\,\mbox{[day]}$ and  $T_2=18.1\,\mbox{[day]}$, respectively. From these values, we obtain the recovery rate from the $i$-th strain class as $\gamma_i=(1-\rho_i)/T_i$, with $i=1,2$. The resulting values are depicted in Table~\ref{Table:01}.

Finally, to estimate the transmission constants $\beta_1$ and $\beta_2$, we analyse the effective reproduction rate (ERR) from the epidemiological data \cite{OurWorldInData}. The largest observed value of the ERR in Chile was the first measure, $R_{0,\,2}^{\mbox{\tiny max}}=2.91$, reported on March 16, 2020. On January 13, 2022, during the rising of the Omicron variant in Chile, the ERR reached the local maximum $R_{0,\,1}^{\mbox{\tiny max}}=1.81$. Following Eqs.~\eqref{Eq:12} and \eqref{Eq:15}, we compute the transmission constants as $\beta_i=R_{0,\,i}^{\mbox{\tiny max}}\delta_i$ with $i=1,2$.








%

\end{document}